\documentclass[aps,prb,preprint,groupedaddress]{revtex4}

\usepackage{graphicx}

\begin{document}

\title{Tight binding description of the electronic response of a
molecular device to an applied voltage}

\author{C. Krzeminski, C. Delerue$^{*}$, G. Allan}

\affiliation{Institut d'Electronique et de Micro\'electronique du Nord, D\'epartement Institut Sup\'erieur d'Electronique du Nord, 41 Boulevard Vauban, 59046 Lille C\'edex, France}

\email[*]{Christophe.Delerue@isen.fr}

\vspace{10cm}

\begin{abstract}
We analyze the effect of an external electric field on the electronic structure of molecules which have been recently studied as molecular wires or diodes. We use a self-consistent tight binding echnique which provides results in good agreement with {\it ab initio} calculations and which may be applied to a large number of molecules. The voltage dependence of the molecular levels is mainly linear with slopes intimately related to the electronic structure of the molecules. We emphasize that the response to the applied voltage is an important feature which governs the behavior of a molecular device.
\end{abstract}


\maketitle

\section{Introduction}

Recent measurements of single molecule transport properties\cite{Aviram98,Reed97,Ohnishi98} represent important contributions to the eventual realization of molecular electronics. This is accompanied by an increasing theoretical effort to understand the relationship between the electronic structure of the molecules and the current-voltage $I(V)$ characteristics. Until recently, most of the calculations based on tight binding (TB) or {\it ab initio} methods \cite{Aviram98} did not explicitly include the effect of the applied voltage on the electronic structure of the molecules. This is a severe drawback since several volts can be applied to the electrodes in contact with a single molecule. In the some cases \cite{Xue99}, the energy of the molecular electronic levels is approximated by a linear function of the voltage but its slope is not calculated or is approximated by 1/2. An approach to overcome these limitations has been proposed\cite{Treboux98,Treboux00} recently on the basis of a Hubbard Hamiltonian \cite{Hubbard63} and has been applied to the azulene molecule \cite{Treboux98} and to polyacene wires \cite{Treboux00}. But the Hubbard Hamiltonian is a model Hamiltonian which only gives qualitative predictions. On the other hand, {\it ab initio} methods cannot be easily applied to complex systems. Thus there is a need for simpler methods which correctly predict the response of the molecules and the dependence of the electronic levels as function of the applied bias. This is a prerequisite to calculate the $I(V)$ characteristics of a molecular device. In this paper, we present self-consistent TB calculations of various molecules. We study the evolution of the dipole and of the electronic levels with the voltage. We obtain results in good agreement with {\it ab initio} calculations in the local density approximation (LDA). This is a good test of our method and allows to extend it to large systems. We conclude that our self-consistent TB technique is a promising approach to study the transport properties in a molecular device and that it may be applied to a wide range of systems.

\section{Calculations}
The molecules considered in this work are shown in Figure \ref{fig:fig1}. In order to compare with previous results \cite{Treboux98,Treboux00}, we study tetracene and azulene molecules as prototypes of molecular wires. The calculations are also applied to quinolinium tricyanoquinodimethanide (Q-3CNQ) and to 3,5-dinitrobenzyl 7-(1-oxohexylamino)-pyren-2-ylcarbamate (OHAPy-C-DNB) which are respectively D-$\pi$-A and D-$\sigma$-A molecules. D and A are respectively electron donor and acceptor. $\pi$ and $\sigma$ are respectively "pi" and "sigma" bridges. These two molecules are intensively studied to make molecular diodes \cite{Ashwell90,Metzger97}. The electronic structure of all the molecules is obtained in LDA and in TB. In LDA, we use the DMOL code \cite{Cerius97} with a double numerical basis set (two atomic orbitals for each occupied orbital in the free atom) together with polarization functions ($2p$ for H, $3d$ for N, O and C). The exchange-correlation energy is approximated by the density functional of ref \cite{Vosko80}. The self-consistent TB technique is presented in another publication \cite{Krzeminski99} where we calculate the electronic structure of thienylenevinylene oligomers, showing a good agreement with LDA calculations and experiments. C, N, O atoms are represented by one $s$ and three $p$ atomic orbitals, and H atoms by one $s$ orbital. The non-diagonal terms of the Hamiltonian matrix H are restricted to first nearest-neighbor interactions and to two-center integrals. They depend on the interatomic distance following Harrison's rules \cite{Harrison80}. The self-consistency is incorporated in the diagonal terms $H_{i \alpha,i \alpha}$ ($\alpha$ is the orbital index, $i$ is the atomic index)

\begin{equation}
H_{i \alpha,i \alpha}=H^{0}_{i \alpha,i \alpha}-\sum_{j}\frac{Q_{j}}{\sqrt{R_{ij}^{2}+(\frac{e^{2}}{U_{0}^{2}})}}-eV_{ext}(i)
\label{eq:eq_one}
\end{equation}

where $H^{0}_{i \alpha,i \alpha}$ define the atomic levels, $U$ is the intra-atomic Coulomb energy, R$_{ij}$ is the distance between atoms $i$ and $j$, and Q$_{j}$ is the net charge on the atom $j$. The TB parameters are determined by a fit of the LDA electronic structure of simple molecules \cite{Krzeminski99}. The parameters for C-C and C-H interactions are given in Ref \cite{Krzeminski99}, those for C-N, N-H, C-O, N-O interactions are in Table \ref{table:table1}. The selfconsitency is obtained with an usual iterative method which is converged when the atomic charges Q$_{j}$ in eq. (\ref{eq:eq_one}) correspond to the charges calculated from the eigenstates of the Hamiltonian. V$_{ext}$ in eq. (\ref{eq:eq_one}) is the electrostatic potential resulting from the applied electrostatic field which, for simplicity, is assumed homogeneous, corresponding to a situation where the electrodes would be far from the molecule. Obviously, to calculate the $I(V)$ curve, we should consider the chemical and electrostatic interactions between the molecule and the electrodes. This can be done in TB but this is beyond the scope of the present paper.

\section{Results and discussions}
\begin{figure}[]
\includegraphics[scale=1.4]{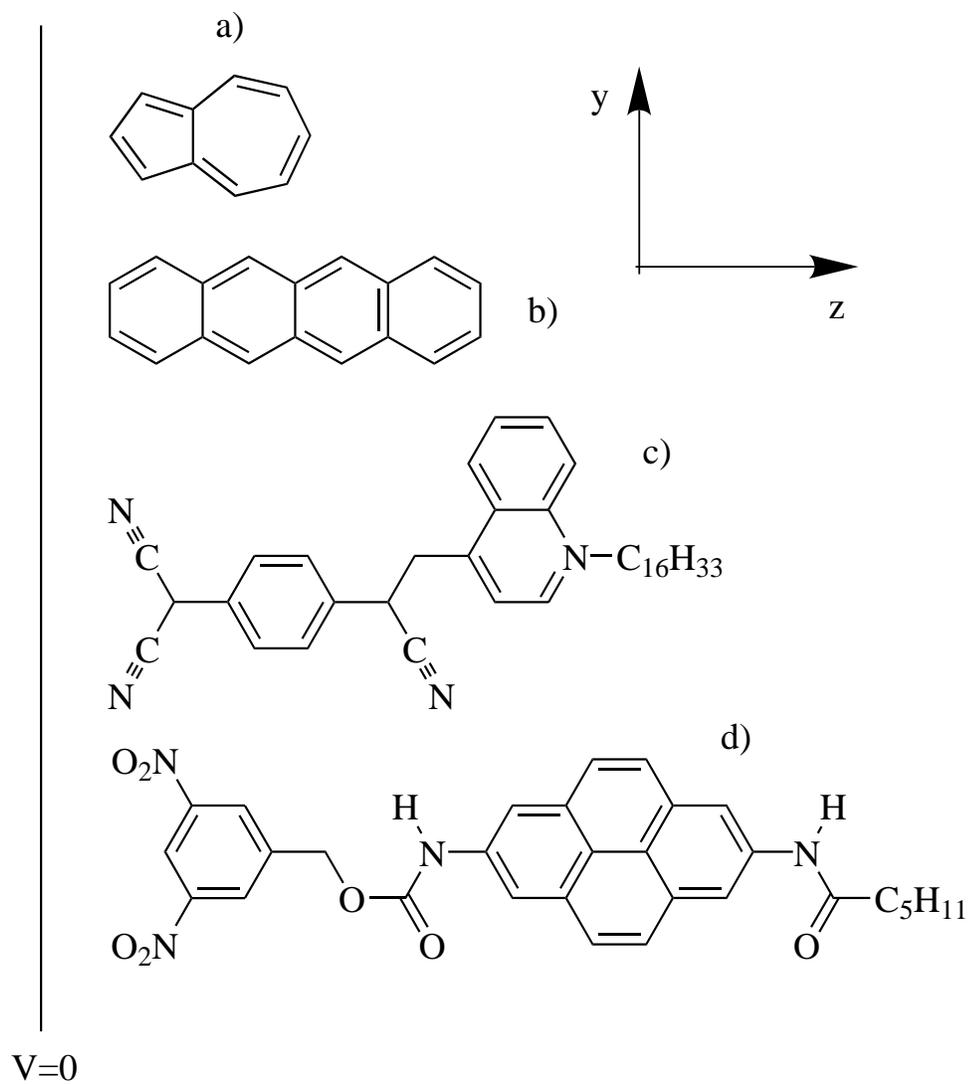}

\caption{\label{fig:fig1} Molecular structure of a) azulene, b) tetracene, c) C$_{16}$H$_{33}$Q-3CNQ, d) OHAPy-C-DNB. The line at the left side of the molecules defines the zero of the potential V$_{ext}$.}
\end{figure}

The applied electric field $E$ is parallel to the long axis $(z)$ of the molecules (Figure \ref{fig:fig1}). The origin of the electrostatic potential (V$_{ext}$ = 0 ) is defined at 2 \AA \ from the left side of the molecules. Thus $V_{ext}(i)= E z_{i}$ where $z_{i}$ is the coordinate of the atom $i$ along the axis $z$. All the results presented in this paper are plotted as a function of the electrostatic potential at 2 \AA \ from the right side of the molecules. Equivalently, a bias of 1 V corresponds to an applied electric field of 88 mV/ \AA, 64 mV/\AA, 57 mV/\AA  \ and 46 mV/\AA \ respectively for azulene, tetracene, Q-3CNQ and OHAPy-C-DNB (the inter-electrode distance is 11.4 \AA, 15.5 \AA, 17.4 \AA \ and 21.6 \AA, respectively). We plot in Figure 2 the voltage dependence of the dipole moment of the four molecules. Azulene has a small permanent dipole moment resulting from its nonalternant character \cite{March92}. Q-3CNQ is characterized by a large dipole moment due to charge transfer between acceptor and donor sites. OHAPy-C-DNB have weaker acceptor and donor characters. We see that the agreement with LDA is good. In particular, the overall magnitude and the slopes of the curves (or the polarizabilities) agree well. It means that the charge transfers induced by the applied bias are correctly described. In ref \cite{Treboux98}, the dipole moment of azulene was not calculated, but equivalently the charge on the five-membered ring was plotted as a function of the voltage. Between -3 V and +3 V, a variation of this charge of 0.5$e$ is reported to be compared with our much smaller value of 0.3$e$. We present in Figure \ref{fig:fig3} the voltage dependence of the energy levels of tetracene. A very good agreement is obtained with LDA results, which was not the case with the Hubbard Hamiltonian \cite{Hubbard63}. The variation of the levels is mainly linear, with a slope 1/2 as expected from the symmetry of the molecule. In spite of an asymmetric atomic structure, quite similar results are obtained for Q-3CNQ (Figure \ref{fig:fig4}) with a slope 0.52 for the highest occupied orbital (HOMO) level and 0.49 for the lowest unoccupied orbital (LUMO) level in TB (resp. 0.56 and 0.55 in LDA). This is a consequence of the delocalization of these orbitals over the whole molecule, which results from an efficient coupling between acceptor and donor sites through the $\pi$ bridge. The situation is completely different in the case of OHAPy-C-DNB (Figure \ref{fig:fig5}) because the $\sigma$ bridge leads to a much smaller coupling between the acceptor and donor parts. Therefore, the HOMO is mainly localized on the acceptor site and the LUMO on the donor site. As the donor site is closer to $z = 0$ where the origin of the potential is defined, the voltage dependence is smaller for the HOMO than for the LUMO. All these important tendancies are predicted similarly in LDA and TB, even if TB calculations underestimate the HOMO-LUMO gap compared to LDA in the case of OHAPy-C-DNB. Thus, even if Q-3CNQ and OHAPy-C-DNB are donor-acceptor molecules, their electrical characteristics will completely differ because of different voltage dependence of the levels. We have also studied the effect of an applied voltage on other molecules like thienylenevinylene oligomers \cite{Krzeminski99} and the $\gamma$-hexadecylquinolinium tricyanoquinodimethanide (C$_{16}$H$_{33}$Q-3CNQ, Figure \ref{fig:fig1}). The response is also linear between -3V and +3V, and the agreement between TB and LDA remains good. The case of C$_{16}$H$_{33}$Q-3CNQ is interesting because it differs from Q-3CNQ only by the replacement of an hydrogen atom by an hexadecyl group. Thus the HOMO and the LUMO are almost the same in the two molecules. However, their voltage dependence in C$_{16}$H$_{33}$Q-3CNQ is characterized by a much smaller slope ($\sim 0.2$) due to the fact that a large part of the voltage drop takes place in the long C$_{16}$H$_{33}$ group which is weakly polarizable, while the HOMO and the LUMO are localized in the remaining part of the molecule. Once again, we expect completely different $I(V)$ curves for these two molecules in spite of their important similarities.

\section{Conclusion}
We have presented a self-consistent TB method to study the response of molecules to an electric field. Our work justifies for a broad range of molecules the hypothesis of a linear dependence of the molecular levels with respect to an applied voltage \cite{Xue99}, even up to several volts. The self-consistent TB method provides a simple approach to calculate this dependence with enough accuracy. It represents a good compromise between model Hamiltonians and ab-initio calculations. It will be useful to simulate more complex systems, to calculate the charging energy of molecules \cite{Krzeminski99} and to predict the $I(V)$ characteristics of molecular devices.

\begin{acknowledgments}
The Institut d'Electronique et de Micro\'electronique du Nord is UMR 8520 of CNRS.

\end{acknowledgments}

\newpage
%

\begin{table}
\begin{tabular}{|cccccc|}
\hline
Interactions&C-O  &N-H  &N-O  &C-N  &O-H\\
\hline
(ss $\sigma$)  &-3.62&-7.54&-3.32&-2.90&-6.96\\
(sp $\sigma$)  &+4.96&6.03 &+4.50&+3.68&3.66\\
(pp $\sigma$)  &+5.31&&+4.55&+5.23&\\
(pp $\pi$)     &-2.00&&-1.98&-2.00&\\
d$_{0}$          &1.54&1.07&1.54&&1.07\\
\hline
Atomic levels  &N&O&&&\\
E$_{s}$&-12.17&-16.17&&&\\
E$_{p}$&-7.97&-8.77  &&&\\
\hline
\end{tabular}
\caption{\label{table:table1}TB parameters in electron volts. The first nearest-neighbor interactions are given for an interatomic distance d$_{0}$ (\AA)}
\end{table}

\vspace{5cm}
\newpage
%

\begin{figure}[]
\includegraphics[scale=0.7]{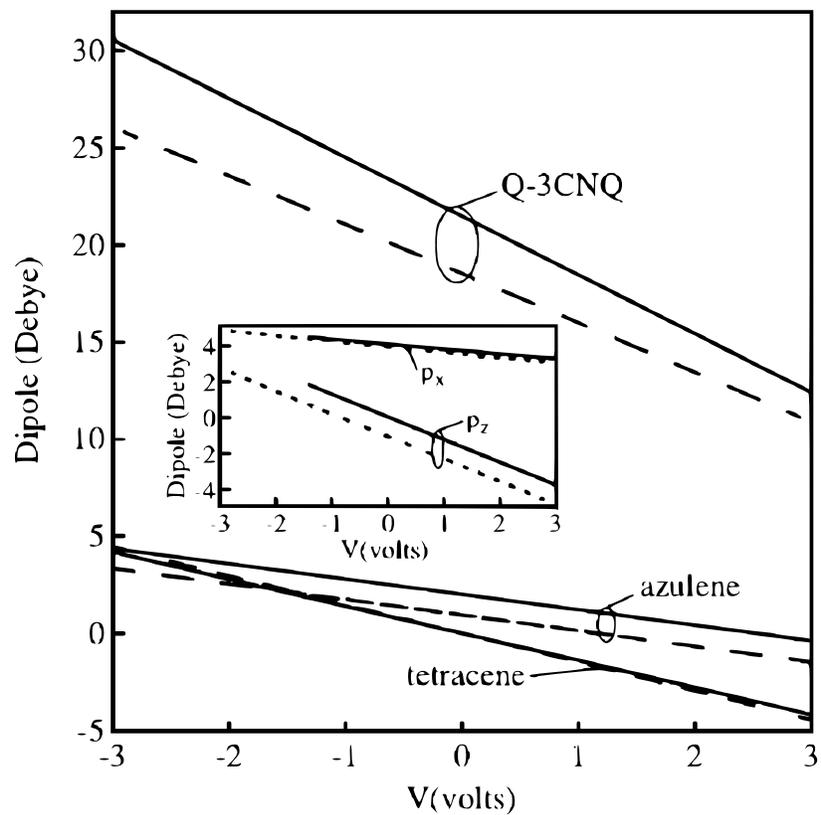}

\caption{\label{fig:fig2}Voltage dependence of the dipole moments along the axis z (straight lines: TB; dashed lines: LDA). Inset: dipole moments along z and x axes for OHAPy-C-DNB.}
\end{figure}

\begin{figure}[]
\includegraphics[scale=0.8]{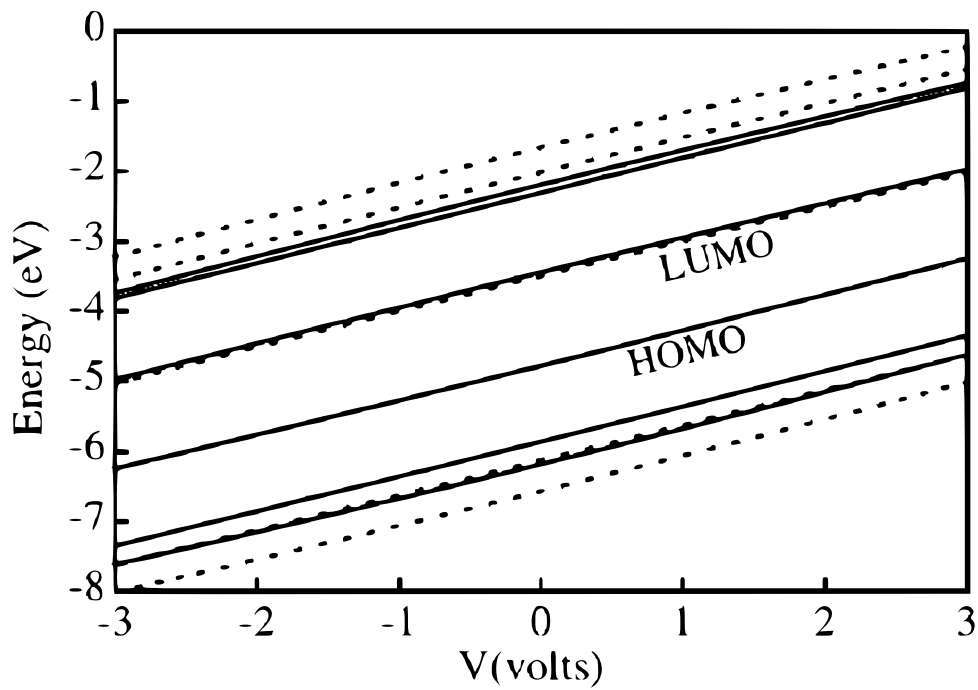}

\caption{\label{fig:fig3}Voltage dependence of the energy levels of tetracene (straight lines: TB; dashed lines: LDA).}
\end{figure}

\begin{figure}[]
\includegraphics[scale=0.8]{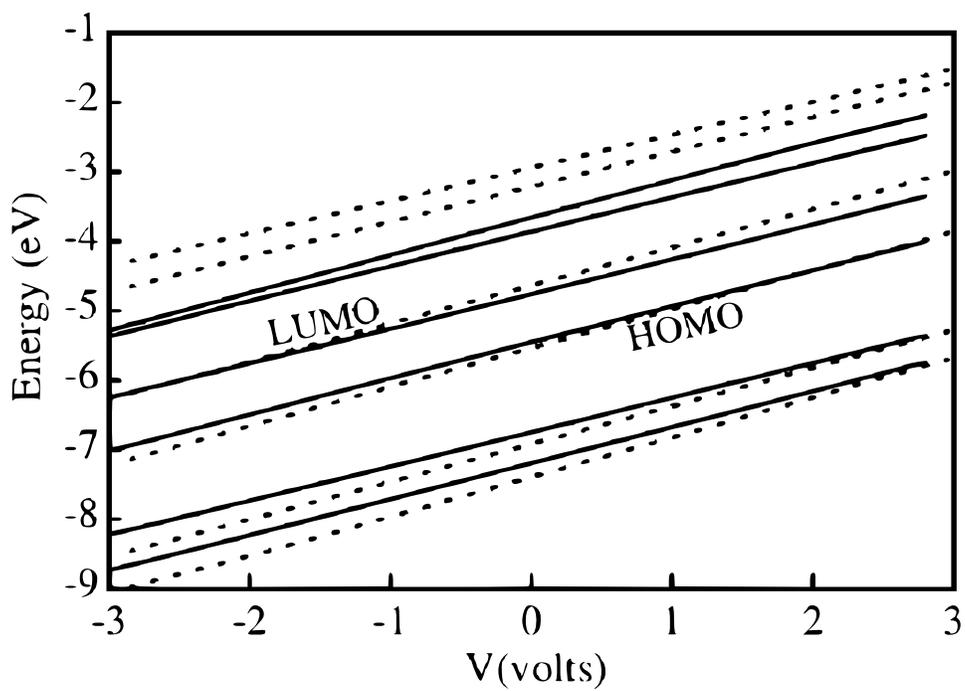}
\caption{ \label{fig:fig4} Voltage dependence of the energy levels of Q-3CNQ (straight lines: TB; dashed lines: LDA).}
\end{figure}

\begin{figure}[]
\includegraphics[scale=0.7]{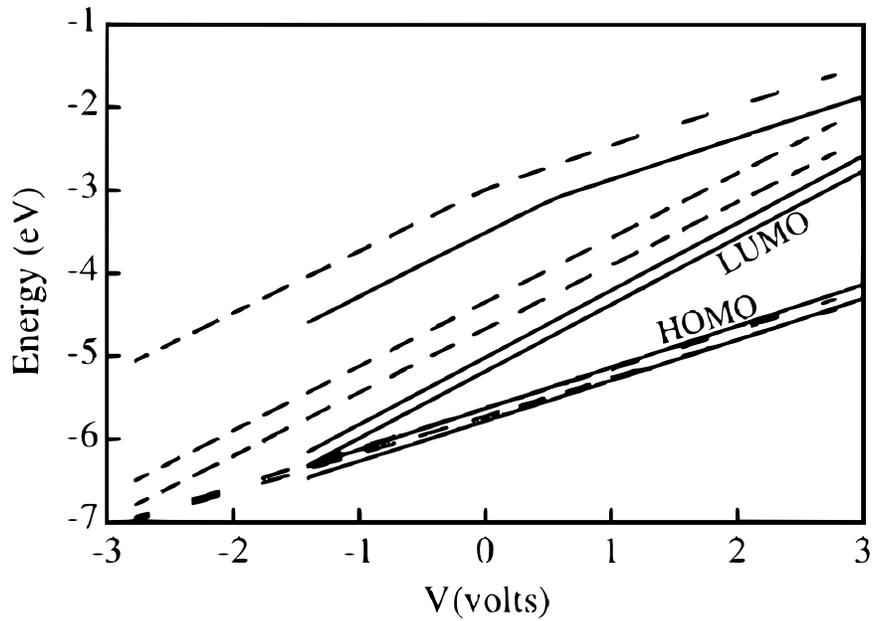}
\caption{\label{fig:fig5} Voltage dependence of the energy levels of OHAPy-C-DNB (straight lines: TB; dashed lines: LDA).}
\end{figure}

\end{document}